\documentclass[aip,jap,amsmath,amssymb,onecolumn,notitlepage]{revtex4-1}

\pdfoutput=1

\usepackage{amssymb,graphicx,lineno}
\usepackage{rotating}
\usepackage{enumerate}
\usepackage{setspace}
\usepackage{multirow, color}
\usepackage[colorlinks=true,allcolors=blue]{hyperref}
\usepackage{siunitx}
\usepackage{subcaption}
\usepackage{tabularx}
\usepackage{booktabs,threeparttable}
\usepackage{adjustbox}

\begin{document}

\newcommand{\stgb}[5]{\ensuremath{\Sigma#1\left(#2#3#4\right)\theta=#5^\circ}} 
\newcommand{\atgb}[8]{\ensuremath{\Sigma#1\left(#2#3#4\right)_1/\left(#5#6#7\right)_2\Phi=#8^\circ}} 
\newcommand{\planef}[3]{\ensuremath{\left\{ #1#2#3 \right\} }}
\newcommand{\plane}[3]{\ensuremath{\left( #1#2#3 \right)}}
\newcommand{\dirf}[3]{\ensuremath{\left\langle #1#2#3 \right\rangle}}
\newcommand{\dir}[3]{\ensuremath{\left[ #1#2#3 \right]}}

\title[He$_n$V in Fe Grain Boundaries]{Binding of He$_n$V Clusters to $\alpha$-Fe Grain Boundaries }

\author{M.A. Tschopp}
\email[Corresponding author, email: ]{mark.tschopp@gatech.edu}
\address{U.S. Army Research Laboratory, Aberdeen Proving Ground, MD 21005}
\address{Center for Advanced Vehicular Systems, Mississippi State University, Starkville, MS 39759}
\author{F. Gao}
\address{Pacific Northwest National Laboratory, Richland, WA 99352}
\author{K.N. Solanki}
\address{Arizona State University, Tempe, AZ 85287}

\begin{abstract}
The objective of this research is to explore the formation/binding energetics and length scales associated with the interaction between He$_n$V clusters and grain boundaries in bcc $\alpha$-Fe.  In this work, we calculated formation/binding energies for 1--8 He atoms in a monovacancy at all potential grain boundary sites within 15 \AA\ of the ten grain boundaries selected (122106 simulations total).  The present results provide detailed information about the interaction energies and length scales of 1--8 He atoms with grain boundaries for the structures examined.  A number of interesting new findings emerge from the present study.  First, the $\Sigma3$\plane112 `twin' GB has significantly lower binding energies for all He$_n$V clusters than all other boundaries in this study.  For all grain boundary sites, the effect of the local environment surrounding each site on the He$_n$V formation and binding energies decreases with an increasing number of He atoms in the He$_n$V cluster.  Based on the calculated dataset, we formulated a model to capture the evolution of the formation and binding energy of He$_n$V clusters as a function of distance from the GB center, utilizing only constants related to the maximum binding energy and the length scale.  

\end{abstract}

\maketitle

\section{\label{sec:sec1}Introduction}
The ability to predict the mechanical behavior of current and future nuclear power reactors necessitates understanding the atomic interactions associated both with radiation damage phenomena and grain boundaries in polycrystalline nuclear materials \citep{Ull1984}.  In particular, future fusion reactors will produce a much larger amount of both He and H as compared to fission reactors, hence the microstructure of the structural materials used in fusion reactors will be much more sensitive to interactions with He defects \cite{Blo2007,Zin2009}.  In terms of radiation damage, the production of helium through (n,$\alpha$) transmutation reactions causes both microstructure evolution and drastic property changes in the first-wall and blanket structural materials of fusion reactors.  The production of single helium atoms and small He clusters in the metal lattice is inherently a problem that occurs at the nanoscale.  The subsequent diffusion of He and He clusters results in the nucleation and growth of He bubbles on grain boundaries and within the lattice, which lead to a macroscopic deterioration of material properties including void swelling, surface roughening and blistering, and high temperature intergranular embrittlement \citep[\textit{e.g.},][]{Ull1984,Blo2007,Zin2009,Yam2006,Tri2003,Man1983,Sto1990,Vas1991,Sch2000}.  While the production and diffusion of He occurs at the nanoscale, these other processes develop at larger length scales over long time scales, which necessitates developing predictive multiscale models for material behavior under irradiation conditions that couples multiple simulation methods at different length and time scales.   Developing this predictive capability will require an understanding of the mechanisms associated with radiation damage phenomena, of the He interaction with microstructures, and of the associated uncertainties.  

It is well known that He interactions in Fe play an important role in the mechanical behavior of steel alloys.  There have been a number of quantum mechanics and molecular dynamics simulations that have examined how He and He clusters affect single crystal lattice properties and physical properties in $\alpha$-Fe \citep{Fu2005, Fu2007, Sel2006, Mor2003a,Mor2003c,Gao2011,Hei2006,Hei2007a,Ter2009, Yan2013, Yan2008b, Yan2008c, Yan2007a, Zu2009, Ven2006, Ste2010,Ste2011, Hay2012, Jus2009}.  For instance, density functional theory (DFT) simulations have been used to show that interstitial He atoms strongly interact with vacancies and can also be trapped by interstitial atoms (binding energy of 0.3 eV) \cite{Fu2005}.  Ventelon, Wirth, and Domain \cite{Ven2006} probed the interactions between He and self-interstitial atoms (SIAs) in $\alpha$-Fe and found strong binding behavior between interstitial He and SIA clusters, which corresponded with the SIA defect strain field.  Other atomistic studies have examined how He and H interact within the single crystal lattice to form complex He--H clusters \cite{Hay2012} or how He impacts the production of  irradiation-induced point defects in an Fe--Cr matrix \cite{Jus2009}.  Stewart \textit{et al.} \citep{Ste2010,Ste2011} recently used several Fe-He potentials \citep{Jus2008,Sel2007,Wil1972} to show the effect of the interatomic potential on the resulting dynamics of He transport and He clustering in Fe.  Ascertaining the reactions that occur and quantifying their energetics are very important for a fundamental understanding of how point defects, impurities, substitutional atoms, and helium atoms interact in the single crystal lattice of $\alpha$-Fe.  Furthermore, this information is useful for models that explore the kinetics of He diffusion, trapping (clustering), and detrapping (emission), such as rate theory models \cite{Ort2007,Ort2009,Ort2009a, Xu2010}, kinetic Monte Carlo models \cite{Deo2007a,Deo2007b}, and/or phase field models \cite{Hu2009,Zha2012a}.  

The grain boundary itself and its atomic configuration within these alloy systems plays a significant role in trapping point defects and various atomic species.  There have been a number of recent studies using both first principles and molecular dynamic simulations that have examined how solutes and impurities segregate to grain boundaries within bcc metals \cite{Jani2003,Yama2007,Kiej2008,Jani2008,Osso2009,Wach2011,Sol2013,Rho2013}.  Despite this fact, there have been relatively few studies that have focused on He interactions with grain boundaries \cite{Kur2004,Gao2006,Kur2008, Gao2009,Ter2010b,Zha2010i,Zha2012z,Ter2011, Zha2013, Zha2013z, Tsc2014}.  These prior works have been significant for understanding the migration paths and mechanisms of He for a few boundaries using the dimer method \cite{Gao2006, Gao2009}, understanding migration of interstitial He in different grain boundaries using molecular dynamics \cite{Ter2010b}, understanding how the grain boundary strength is affected by He \cite{Zha2010i,Zha2012z} or He bubbles \cite{Ter2011}, or understanding the diffusion and stability of He defects in grain boundaries using first principles \cite{Zha2013, Zha2013z}.  For instance, Kurtz and Heinisch \cite{Kur2004} used a Finnis--Sinclair potential (detailed in Morishita et al.~\cite{Mor2003c}) to show that interstitial He was more strongly bound to the grain boundary core than substitutional He.  Kurtz and Heinisch also found that the maximum He binding energy increases linearly with the grain boundary excess free volume, similar to prior work in fcc nickel \cite{Bas1985}.  In subsequent studies, Gao, Heinisch, and Kurtz \cite{Gao2006} found a relationship between the maximum binding energy and grain boundary energy as well.  Additionally, Gao et al.~started to detail the diffusion trajectories of interstitial and substitutional He atoms along a $\Sigma3$ and $\Sigma11$ grain boundary and found that the dimensionality of migration of interstitial He may depend on temperature (e.g., in the $\Sigma3$\plane112 boundary).  Some recent work has utilized first principles to quantify binding strengths of He and He-vacancy clusters at the $\Sigma5$\plane310 symmetric tilt grain boundary \cite{Zha2013z}.  There are still a number of unresolved issues relating to how He interacts with grain boundaries, though.  For instance, these studies often focus on one He atom in interstitial or substitutional sites, but often do not extend to multiple He atoms interacting with grain boundary sites.  Also, atomistic studies often have not examined a wide range of grain boundary structures to understand the influence of macroscopic variables on He interactions.  Moreover, while the highly non-uniform He binding energies in the grain boundary core have been previously pointed out \cite{Kur2004}, relating these to per-atom metrics based on the grain boundary local environment has not been pursued as frequently.  Additionally, while there is an increasing awareness of interatomic potential effects, many of the interatomic potentials used previously for some of these grain boundary studies have been improved upon with updated interatomic potential formulations \cite{Gao2011, Sel2007, Sto2010} and/or more recent quantum mechanics results showing how magnetism affects He defects in $\alpha$-Fe \cite{Sel2005, Sel2006, Zu2009}.  

In the present work, we have focused on how the local grain boundary structure interacts with He atoms and how the local atomic environment at the boundary influences the binding energetics of He$_n$V clusters with up to 8 He atoms inside a monovacancy.  Recently, Tschopp and colleagues utilized an iterative approach to systematically quantify the interactions between point defects, carbon, hydrogen, and helium with Fe grain boundaries \cite{Tsc2011,Tsc2012a,Rho2013, Sol2013,Tsc2014}.  Herein, this approach is applied along with using multiple different starting positions, or instantiations, about each site to more precisely probe the formation and binding energy landscape about ten grain boundaries.  In this paper, we present this approach for eight He$_n$V clusters and we explore how the grain boundary structure interacts with He$_n$V clusters using molecular static calculations.  Moreover, we have also explored how different per-atom local environment metrics compare with the calculated energies for the different He$_n$V clusters and the energetics of incorporating additional interstitial He atoms into the He$_n$V clusters.  A number of interesting new findings emerge from the present study.  First, the $\Sigma3$\plane112 `twin' GB has significantly lower binding energies for all He$_n$V clusters than all other boundaries in this study.  For all grain boundary sites, the effect of the local environment surrounding each site on the He$_n$V formation and binding energies decreases with an increasing number of He atoms in the He$_n$V cluster.  Based on the calculated dataset, we formulated a model to capture the evolution of the formation and binding energy of He$_n$V clusters as a function of distance from the GB center, utilizing only constants related to the maximum binding energy and the length scale.  This work significantly enhances our understanding of the energetics involved with how the grain boundary structure interacts with He$_n$V clusters and how ultimately this may affect He (re-)combination and embrittlement near grain boundaries in polycrystalline steels.

\vspace{-10pt}
\section{\label{sec:sec2}Methodology}

\vspace{-10pt}
\subsection{Grain Boundaries}
The interaction between helium--vacancy clusters and iron grain boundaries was investigated by using ten different grain boundaries and multiple different He$_n$V clusters ($n=1$--$8$)  for multiple sites (866 total sites) within 15 \AA\ of the boundary (122106 simulations total).  Table \ref{table1} lists the ten grain boundaries studied, their dimensions in terms of lattice units, the number of atoms and the interfacial energy.  These grain boundaries represent the ten low coincident site lattice (CSL) boundaries ($\Sigma\leq13$) within the \dirf100 and \dirf110 symmetric tilt grain boundary (STGB) systems.  This is a subset of those boundaries used in prior studies of point defect absorption (vacancies and self-interstitial atoms) by a large range of grain boundary structures in pure $\alpha$-Fe \cite{Tsc2011, Tsc2012a} and is identical to those used in our previous study of 1--2 atom He defect interactions with grain boundaries \cite{Tsc2014}.  

The current set of boundaries includes four \dirf100 STGBs ($\Sigma5$,$\Sigma13$) and six \dirf110 STGBs ($\Sigma3$,$\Sigma9$,$\Sigma11$).  Recent experimental characterization of steels has shown that several of these symmetric tilt grain boundaries are observed at a concentration higher than random grain boundaries \cite{Bel2013a, Bel2013b}.  For example, Beladi and Rollett quantified that the $\Sigma3$\plane112 symmetric tilt grain boundary is observed at $>$10 multiples of a random distribution (MRD) of grain boundaries \cite{Bel2013a, Bel2013b}, i.e., much larger than would be expected.  While the experimental observation of \dirf100 symmetric tilt grain boundaries ($\Sigma5$, $\Sigma13$ GBs) is below 1 MRD, these grain boundaries are commonly used in DFT studies due to the low periodic distances required in the grain boundary plane.  The present set of boundaries is smaller than those previously explored \cite{Tsc2011, Tsc2012a} for two reasons.  First, since we explored multiple starting configurations for the He$_n$V clusters in this study, a larger number of simulations were required for each grain boundary than for the point defect studies, which only considered a single vacancy or self-interstitial atom.  Second, our prior study \cite{Tsc2012a} found that, aside from a few boundaries (e.g., the $\Sigma3$\plane112 STGB, included herein), most grain boundaries had similar characteristics with respect to point defect interactions.  These results suggest that the ten boundaries explored within can supply ample information about the interaction of He defects with low-$\Sigma$ grain boundaries, and perhaps shed insight on general high angle grain boundaries as well.  

The simulation cell consisted of a 3D periodic bicrystalline structure with two periodic grain boundaries, similar to prior grain boundary studies \cite{Rit1996,Spe2007,Tsc2007}.  The two mirror-image grain boundaries are separated by a minimum distance of 12 nm to eliminate any effects on energies due to the presence of the second boundary.  While the grain boundaries were generated using the minimum periodic length in the grain boundary period direction and the grain boundary tilt direction ($x$- and $z$- directions, respectively), it was found that the formation energies for the defects were influenced for periodic lengths below 4$a_0$.  That is, the periodic image of the defect and/or its influence on the surrounding lattice can significantly affect the defect's formation energy.  Hence, multiple replications in the grain boundary tilt direction and the grain boundary period direction were used.  For instance, the final dimensions for the $\Sigma5$\plane210 GB resulted in a vacancy formation energy far away from the boundary that was within 0.015\% of that within a 2000-atom bcc single crystal (i.e., 10$a_0$ per side).  This criteria resulted in simulation cell sizes on the order of 4660--9152 atoms ($\Sigma{13}$\plane510 and $\Sigma{11}$\plane113, respectively).  All of the simulations were performed with a modified version of the MOLDY code \cite{Ref2000,Ack2011,Gao1997}.

\begin{table}
\centering
\caption{\label{table1} Dimensions of the bicrystalline simulation cells used in this work along with the $\Sigma$ value of the boundary, the grain boundary plane (normal to the $y$-direction), the misorientation angle $\theta$ about the corresponding tilt axis ($z$-direction), the grain boundary energy, and the number of atoms.  The cell dimensions were chosen to ensure convergence of the formation and binding energies of the inserted He$_n$V clusters.  }
\begin{scriptsize}
\begin{ruledtabular}
\begin{tabular}{ccccccccc}
Sigma \&  & GB tilt & $\theta$ & GB energy & $x$ & $y$ & $z$ & Number of & Free Volume \\
GB plane &  direction & ($^\circ$) & (\si{mJ.m^{-2}}) & (\si{\angstrom}) & (\si{nm}) & (\si{\angstrom}) & atoms &  (\AA$^3$/\AA$^2$)\\
\hline \\ [-1.5ex]
$\Sigma3$\plane111 & \dirf110 & 109.47$^\circ$ & \SI{1308} & \SI{21.0}  & \SI{24.8}  & \SI{16.2}  & \num{7200} & 0.35 \\
$\Sigma3$\plane112 & \dirf110 & 70.53$^\circ$ & \SI{260} &  \SI{14.8} & \SI{25.2}  &  \SI{16.2} & \num{5184} & 0.01 \\
$\Sigma5$\plane210 & \dirf100 & 53.13$^\circ$ &   \SI{1113} &  \SI{19.2} &  \SI{24.5} & \SI{14.3} & \num{5730} & 0.35 \\
$\Sigma5$\plane310 & \dirf100 & 36.87$^\circ$ &    \SI{1008} & \SI{18.1}  &  \SI{25.3} & \SI{14.3} &\num{5600} & 0.30 \\
$\Sigma9$\plane221 & \dirf110 & 141.06$^\circ$ &    \SI{1172} & \SI{17.1}  &  \SI{24.2} & \SI{16.2} &\num{5728} & 0.19 \\
$\Sigma9$\plane114 & \dirf110 & 38.94$^\circ$ &   \SI{1286} & \SI{24.2}  & \SI{25.5}  & \SI{16.2} &\num{8576} & 0.35 \\
$\Sigma11$\plane113 & \dirf110 & 50.48$^\circ$ & \SI{1113} & \SI{26.8}  & \SI{24.7}  & \SI{16.2}  &  \num{9152} & 0.26 \\
$\Sigma11$\plane332 & \dirf110 & 129.52$^\circ$ & \SI{1020} &  \SI{18.9} & \SI{24.1}  &  \SI{16.2} &  \num{6336} & 0.21 \\
$\Sigma13$\plane510 & \dirf100 & 22.62$^\circ$ & \SI{1005} &  \SI{14.6} & \SI{26.1}  & \SI{14.3}  &  \num{4660} & 0.27 \\
$\Sigma13$\plane320 & \dirf100 & 67.38$^\circ$ & \SI{1108} &  \SI{20.6} & \SI{24.7}  & \SI{14.3}  &  \num{6220} & 0.23 \\
\end{tabular}
\end{ruledtabular}
\end{scriptsize}
\end{table}

Table \ref{table1} also lists several properties of the ten grain boundaries.  First, notice that the grain boundary energies range from 260--1308 \si{mJ.m^{-2}}, although the majority of the CSL boundaries have energies $>$1000 \si{mJ.m^{-2}}.  Also, all boundaries are high angle grain boundaries, based on a 15$^\circ$ Brandon criterion for low/high angle grain boundaries.  Additionally, note that while the misorientation angles $\theta$ refer to the conventional misorientation angle-energy relationships (e.g., in Ref.~\onlinecite{Tsc2012a}), the disorientation angle, or minimum angle to rotate lattice A to lattice B, is the same for the two instances of each $\Sigma$ boundary.  The misorientation angles are based on deviation from the\plane100 planes in the \dirf100 and \dirf110 STGB systems.  The grain boundary energies are similar to those previously calculated (e.g., $\Sigma5$\plane310 and $\Sigma13$\plane320 GBs are within 2\% and 7\%, respectively, of a prior study \cite{Li2013}).  The grain boundary structures vary for the ten grain boundaries.  Further details on the grain boundary structure are given in Tschopp et al.~\cite{Tsc2012a}.  The grain boundary structures have been compared with computed structures using quantum mechanics, when possible.  For instance, Bhattacharya et al.~used DFT to calculate grain boundary structures for $\Sigma3$\plane111 and $\Sigma11$\plane332 GBs \cite{Bha2013}, which are identical to those computed in the present work.  Moreover, the relationship between the grain boundary energies and excess free volume for the $\Sigma3$\plane111 and $\Sigma11$\plane332 GBs also agrees with previous studies \cite{Shi2009,Bha2013,Kur2004}, as well as with other studies that have found that the $\Sigma3$\plane112 GB has a much lower grain boundary energy and excess free volume in comparison to the $\Sigma3$\plane111 GB \cite{Shi2009,Tsc2012a}.  Additionally, the $\Sigma5$\plane310 and $\Sigma9$\plane114 GB structures also agree with previously calculated first principles structures \cite{Zha2013, Zha2013z}.  Also included in this table is the excess free volume, which was calculated using a previous methodology for calculating excess volume \cite{Wol1991, Tsc2007b} whereby the volume occupied by the bicrystal simulation cell is compared to an equivalent volume of a perfect single crystal lattice and divided by the total grain boundary area.  

\vspace{-10pt}
\subsection{Interatomic Potential}
The Fe--He interatomic potential fitted by Gao et al.~\cite{Gao2011} to ab initio calculations using an s-band model was used in the present atomistic modeling.  This interatomic potential is based on the electronic hybridization between Fe $d$-electrons and He $s$-electrons to describe the Fe--He interaction.  The single element potentials utilized in the formulation of this potential are the Ackland and Mendelev (AM) potential for the Fe--Fe interactions \cite{Ack2004} and the Aziz et al.~Hartree--Fock-dispersion pair potential (Aziz-potential) \cite{Azi1995} for the He--He interactions.  The atomic configurations and formation energies of both single He defects (substitutional, tetrahedral, and octahedral He) and small interstitial He clusters (He$_2$V, He$_3$V, and He--He di-interstitial) were utilized in the fitting process.  Calculations using this interatomic potential show that both tetrahedral and octahedral interstitials are stable, with tetrahedral He being the most stable interstitial configuration \cite{Gao2011}, which agrees with previous ab initio calculations \cite{Sel2005, Zu2009}.  The binding properties of the He$_x$V and He$_x$ interstitial clusters are in reasonable agreement with ab initio and previous potential results.  This potential has been previously used to investigate the emission of self-interstitial atoms from small He clusters in the $\alpha$-Fe matrix and to show the dissociation of a di-interstitial He cluster at temperatures $>$400 K.  The aforementioned potential is deemed appropriate for studying the He interaction with grain boundaries in this work.  In addition, the recent first principles calculations of energetic landscape and diffusion of He in $\alpha$-Fe grain boundaries demonstrate that the potentials used in the present study satisfactorily describe the He behavior at the GBs \cite{Zha2013}.  In fact, this study has shown that there is good agreement between the vacancy formation energies for $n$ vacancies ($n\le{4}$) in the $\Sigma3$\plane111 GB between DFT results and the present empirical potential \cite{Zha2013}.  Furthermore, the present interatomic potential has shown good agreement with DFT results for interstitial and substitutional formation energies for multiple layers and multiple boundaries \cite{Zha2013}.  

\begin{table}
\centering
\caption{\label{table2} Formation energies for the Fe--He empirical potential (EP) \cite{Gao2011} used in the present work compared to other empirical potentials \cite{Sel2007,Che2010_jnm} and DFT results  \cite{Sel2005,Sel2006,Sel2007,Fu2007, Zu2009}.}
\begin{scriptsize}
\begin{ruledtabular}
\begin{tabular}{lccccc}
Cluster & $HeV$ & $He_{oct}$  & $He_{tet}$ & $He_{2}V$ & $He_{2}$ \\
\hline \\ [-1.5ex]
VASP \cite{Sel2007} & $4.08\left(3.73\right)$ eV\tnote{a} & $4.60$ eV  & $4.37$ eV & $6.63$ eV & $8.72$ eV \\
VASP \cite{Zu2009} & 4.34 eV & 4.75 eV & 4.40 eV  & -- & -- \\
SIESTA \cite{Fu2007} & 4.22 eV & 4.58 eV & 4.39 eV & -- & -- \\
EP \cite{Che2010_jnm} & 3.87 eV & 4.57 eV & 4.45 eV & -- & -- \\
EP \cite{Sel2007} & 3.75 eV & 4.57 eV & 4.26 eV & 6.46 eV & 9.37 eV \\
EP \cite{Gao2011}  & $3.76$ eV\tnote{b} & $4.47$ eV\tnote{b}  & $4.38$ eV\tnote{b} & $6.87$ eV\tnote{b} & $8.49$ eV\tnote{b} \\  
\end{tabular}
        \begin{tablenotes}
		\scriptsize
            	\item [a] The data in parentheses were adjusted by Seletskaia et al.~\cite{Sel2007} for their empirical potential fitting.
            	\item [b] The calculated He formation energies are in agreement with previous results \cite{Gao2011}.
        \end{tablenotes}
\end{ruledtabular}
\end{scriptsize}
\end{table}

Previous calculations for He defects with 1--2 atoms \cite{Tsc2014} have also been compared to recent DFT formation/binding energies for the $\Sigma5$\plane310 GB \cite{Zha2010i,Zha2013z}.  For instance, Table \ref{table3} compares the minimum formation energy, mean binding energy, and maximum binding energy calculated herein with those from Zhang et al.~\cite{Zha2010i,Zha2013z}.  The trend of the data with respect to the ordering of formation energies for different He defects agrees well between the Fe--He interatomic potential and DFT.  The minimum formation energies for the $\Sigma5$\plane310 GB are within $12.5$\% of each other, with a larger discrepancy in the binding energies.  The largest $E_b^{max}$ deviation is for the He$_1$V cluster ($-43$\%), with the other He defects falling within 30\% of DFT values.  The difference does not appear random; formation energies are consistently higher for the Fe--He potential and binding energies are consistently lower.  Hence, the present Fe--He interatomic potential \cite{Gao2011} qualitatively agrees with previous DFT calculations \cite{Zha2010i,Zha2013z} and both formation/binding energies agree with DFT within the calculated differences. 

\begin{table}
\centering
\caption{\label{table3} Comparison of Fe--He interatomic potential \cite{Gao2011} and DFT \cite{Zha2010i,Zha2013z} formation and binding energies of various He defects in the $\Sigma5$\plane310 GB}
\begin{threeparttable}[b]
\begin{ruledtabular}
\begin{tabular}{cccccccc}
\multirow{3}*{He Atoms} & \multirow{3}*{Model}& \multicolumn{3}{c}{Substitutional He (He$_1$V)} & \multicolumn{3}{c}{Interstitial He} \\
\cline{3-8} \\ [-1.5ex]
&  & $E_f^{min}$ & $E_{b}$ & $E_{b}^{max}$ & $E_f^{min}$ & $E_{b}$ & $E_{b}^{max}$ \\
&  & (eV) & (eV) & (eV) & (eV) & (eV) & (eV) \\
\hline \\ [-1.5ex]
\multirow{2}*{1} &     MD \cite{Gao2011} &   3.08 &   $E_{b}^{mean}=0.33$ &  0.68 &  3.17 &  $E_{b}^{mean}=0.68$ &  1.21 \\ 
 &      DFT \cite{Zha2010i} &   2.93 & $ \left\{0.64, 1.20, 0.58, 0.28\right\}$ &  1.20 &  2.98 &  $ \left\{1.29, 0.71, 1.43\right\}$  &  1.43 \\ 
\hline \\ [-1.5ex]
\multirow{2}*{2} &      MD \cite{Gao2011} &   5.52 &  $E_{b}^{mean}=0.88$ &  1.35 &  6.32 &  $E_{b}^{mean}=1.24$ &  2.17 \\ 
&      DFT \cite{Zha2013z} &   5.08 &  $ \left\{1.88, 1.45\right\}$ &  1.88 &  5.62 &  $ \left\{2.78\right\}$ &  2.78 \\ 
\end{tabular}
\end{ruledtabular}
\end{threeparttable}
\end{table}

\vspace{-10pt}
\subsection{Helium Clusters}

There are eight different He clusters explored in the present study.  These eight He$_n$V clusters correspond to 1--8 He atoms in a monovacancy, which are denoted as He$_1$V, He$_2$V, He$_3$V, \ldots, He$_8$V.  These He$_n$V clusters result from removing an Fe atom and placing either a single He atom or multiple He atoms nearby the now-vacant site.  For a single He atom in a monovacancy (He$_1$V), the He atom was simply added in the exact location of the removed Fe atom.  

\begin{figure}[h!]
  \centering
  \includegraphics[width=\textwidth,angle=0]{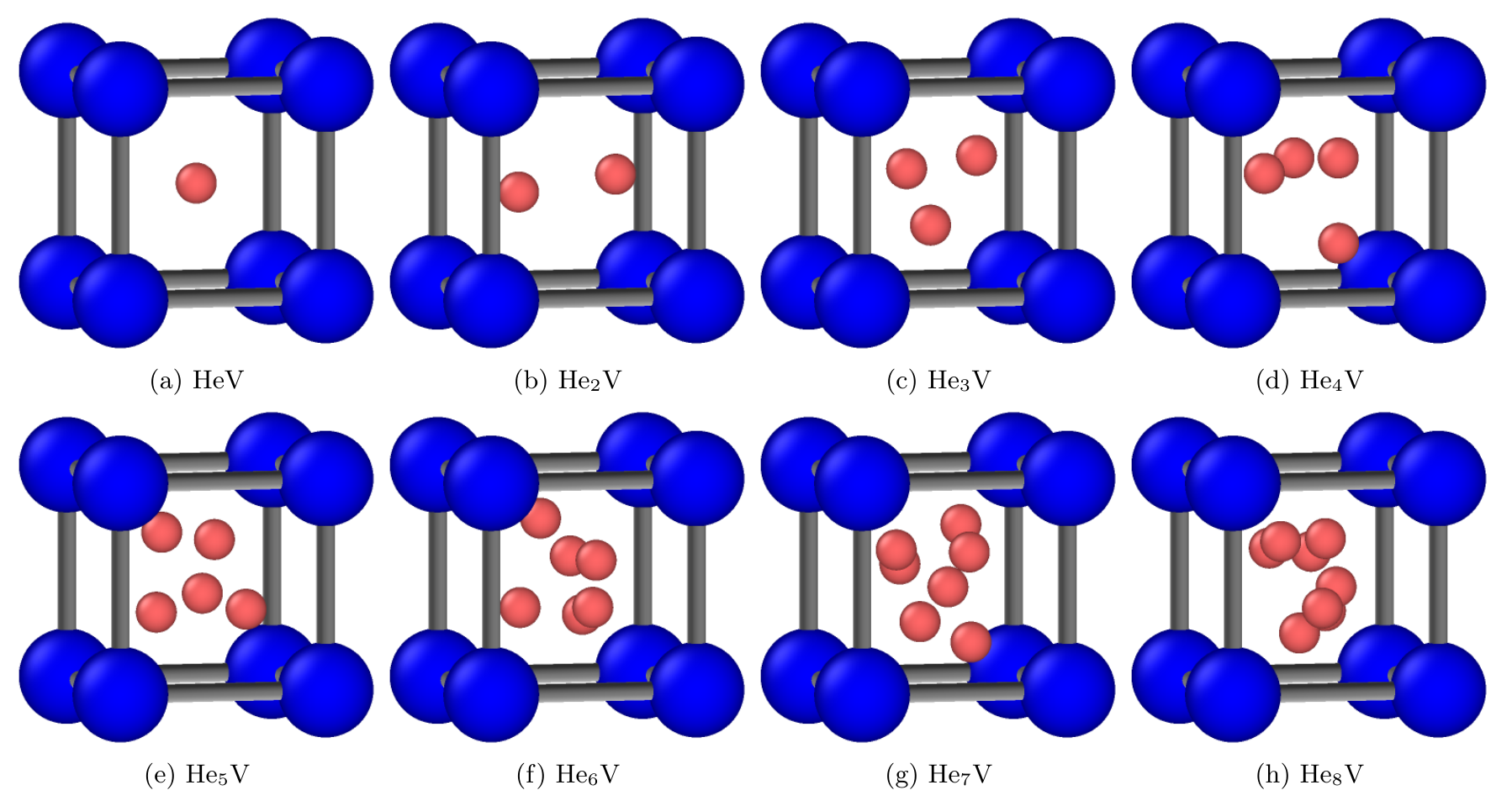}
\caption{ \label{figure1} An example of a starting instantiation (before the energy minimization process) for $n$ He atoms in a monovacancy, i.e., $n=1$--$8$ He (red atoms) in a bcc Fe unit cell (blue atoms) from left to right.  }
\end{figure}

For multiple atoms, a slightly different methodology was used.  In the case of He$_2$V, the two atoms were placed in opposite directions along a randomly-oriented vector eminating from the vacant site with equal distances to the vacant site and a total distance $>$1 \AA.  Since a single instance may not obtain the minimum energy dumbbell, twenty different instances of the starting configurations were used for each potential site for the He$_2$V clusters.  For higher numbers of He atoms, multiple locations were randomly chosen for the He atoms, given the constraint that any two He atoms could not be closer than 1 \AA.  Again, 20 different starting positions were used.  This number of instances (20) was sufficient to obtain a near constant mean formation energy for interstitial He atom (maximum deviation of 0.4\% of bulk value, mean deviation of 0.03\% of bulk value) in the bulk region far away from the grain boundary \cite{Tsc2014}.  Figure \ref{figure1} is an example of one instantiation of the various He$_n$V clusters surrounding the central atom site in the bcc unit cell.   

\vspace{-10pt}
\subsection{Formation and Binding Energies}

The formation energies for the He$_n$V clusters can then be calculated as a function of spatial location of sites and their proximity to the grain boundary.    The formation energy for a He$_n$V cluster containing $n$ He atoms in a monovacancy at site $\alpha$ of a grain boundary configuration is given by
\begin{equation}
  \label{eq:eq1}
	E^{He_nV,\alpha}_{f}=\left(E^{He_nV,\alpha}_{tot} + E_c^{Fe}\right) - \left(E^{GB}_{tot}\right).	
\end{equation}
\noindent Here, $E^{He_nV,\alpha}_{tot}$ is the total energy of the grain boundary configuration with the He$_n$V cluster at site $\alpha$, $E^{GB}_{tot}$ is the total energy of the grain boundary without any defects, and $E_c^{Fe}$ is the cohesive energy of bcc Fe ($E_c^{Fe}=4.013$ eV).  The cohesive energy of He is negligible and not included in Equation \ref{eq:eq1}.  The formation energies of various He$_n$V clusters in the bulk, calculated from atom sites far from the boundary, is given in Table \ref{table4}.  
 
The binding energy of the He$_n$V clusters with the grain boundary is also of interest.  The total binding energy of a He$_n$V cluster interacting with the GB can be directly calculated from the formation energies of the He$_n$V cluster in the bulk and the He$_n$V cluster at the GB.  For instance, the binding energy for a He$_n$V cluster at site $\alpha$ is given by
\begin{equation}
  \label{eq:eq2}
	E^{He_nV^\alpha}_{b}=E^{He_nV, bulk}_{f}-E^{He_nV^\alpha}_{f},	
\end{equation}
\noindent where $E^{He_nV, bulk}_{f}$ and $E^{He_nV^\alpha}_{f}$ are the formation energies of a He$_n$V cluster either in the bulk or at site $\alpha$, respectively.  It can be seen that a positive binding energy represents that it is energetically favorable for the He$_n$V cluster to segregate to the GB, while a negative binding energy represents that the He$_n$V cluster does not want to segregate to the GB.  Table \ref{table4} lists the maximum values of $E_b$ for all ten GBs for each He$_n$V cluster.

\begin{table}
\centering
\caption{\label{table4} Bulk formation energies $E^{He_nV, bulk}_{f}$, bulk binding energies $E^{He_nV, bulk}_{b}$ and the maximum GB binding energies \textrm{max}($E^{He_nV^\alpha}_{b}$) for various He$_n$V clusters (in eV).}
\begin{ruledtabular}
\begin{tabular}{@{}ccccccccc@{}}
Cluster & He$_1$V	& He$_2$V & He$_3$V & He$_4$V & He$_5$V & He$_6$V & He$_7$V & He$_8$V \\ 
\hline \\ [-2ex]
$E^{He_nV, bulk}_{f}$ &   3.760 &  6.870 &  9.892 & 12.868 & 16.038 & 18.974 & 22.114 & 25.269   \\ 
max($E^{He_nV}_{b}$) &  0.752 &  1.694 &  2.437 &  2.993 &  4.043 &  4.829 &  5.953 &  6.905 \\
$E^{He\leadsto He_nV, bulk}_{b}$  & 1.270 &  1.358 &  1.404 &  1.209 &  1.445 &  1.240 &  1.224  & -- \\
\end{tabular}
\end{ruledtabular}
\end{table}

The binding energy of a He$_n$V with an additional He atom is also of interest.  For instance, in this scenario, we would like to know if a He atom in an interstitial location is either attracted to or repelled from the He$_n$V cluster.  This binding energy is given by 
\begin{equation}
  \label{eq:eq3}
	E^{He\leadsto He_nV^\alpha}_{b}=\left(E^{He_nV^\alpha}_{f}+E^{He}_{f}\right) - E^{He_{n+1}V^\alpha}_{f},	
\end{equation}
\noindent where $E^{He}_{f}$ is the formation energy of an interstitial He atom in the bulk ($E_f^{He}=4.38$ eV).  Table \ref{table4} lists the binding energy of attracting an additional He to the He$_n$V cluster in the bulk, e.g., an interstitial He binding to a He$_2$V cluster to produce a He$_3$V cluster will decrease the system energy by 1.358 eV.  Adding an interstitial He atom to a monovacancy to make He$_1$V is slightly different from Equation \ref{eq:eq3}, but uses the formation energy of a vacancy ($E_f^v=1.72$ eV in bulk) at site $\alpha$ instead.

In this work, for He$_n$V clusters ($n\ge2$), the formation and binding energies for a He$_n$V cluster at site $\alpha$ are the `mean' values from the twenty different instantiations, i.e., 
\begin{equation}
  \label{eq:eq4}
	\bar{E}^{He_nV^\alpha}_{b}=\frac{1}{20}\sum_{1}^{20}E^{He_nV^\alpha}_{b},	
\end{equation}
Previous work found that the mean of the formation/binding energy was a better metric for capturing the variation in these quantities as a function of spatial position than the standard deviation or the extreme values (minimum $E_f$ or maximum $E_b$).  While individually these three measures give information of the distribution of binding and formation energies about each particular site, the remainder of the analysis will focus on the mean formation energies and binding energies of the 20 different instantiations, which is more sensitive to local variations than the maximum binding energy and is more applicable to the energetic favorability of He$_n$V clusters than is the standard deviation.      

\section{Results and Discussion}
\vspace{-10pt}
\subsection{Spatial distribution of binding energies}

\begin{figure}[b!]
  \centering
  \includegraphics[height=65mm,angle=0]{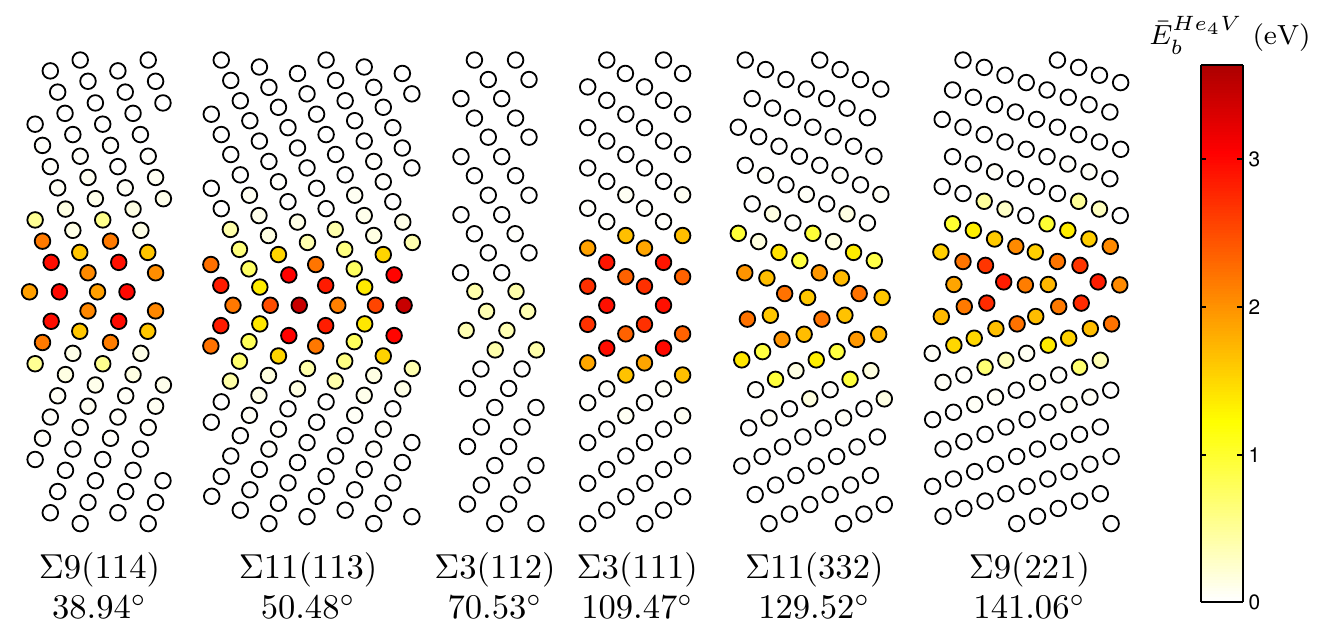}\\
\caption{ \label{figure2} The binding energies of He$_4$V clusters in various sites for the six \dirf110 symmetric tilt grain boundaries.  }
\end{figure}

Figure \ref{figure2} displays the spatial distribution of binding energies for He$_4$V clusters in the \dirf110 symmetric tilt grain boundary system.  Recall that the binding energies correspond to the mean of the binding energies for all 20 instantiations.  The grain boundaries in this figure are arranged in order of increasing misorientation angle.  In Figure \ref{figure2}, each atom represents a site where a He$_4$V cluster was placed and the binding energy was calculated (i.e., each atom is a different simulation).    The minimum periodic length for each grain boundary is shown along the horizontal axis and the length from top to bottom is 30 \AA, with the grain boundary plane centered in the vertical direction.  The binding energy scale shown to the right of Figure \ref{figure2} ranges from 0 eV (bulk lattice) to the maximum calculated binding energy (3.6385 eV for the $\Sigma13$\plane320 GB) from all ten boundaries.  The atoms far away from the boundary are white (binding energy of 0 eV), indicating that there is no energy difference over the bulk lattice.  As the He$_4$V cluster gets closer to the grain boundary, there is a GB-affected region with an increased binding energy for the He$_n$V clusters.  The largest binding energies tend to be along the center of the grain boundary plane or along the 1$^{st}$ layer from the GB plane.  Interestingly, there is a range of binding energies along the central GB plane, as evidenced in the $\Sigma9$\plane114 GB, which contains He$_4$V cluster positions with both the largest binding energy and a binding energy similar to bulk within the central grain boundary plane.  Furthermore, there is a noticeable symmetry to the binding energies about the grain boundary plane due to the symmetric nature of these grain boundary structure.   In terms of variations due to GB structure, the $\Sigma3$\plane112 `twin' grain boundary has a much smaller binding energy in the GB region than all other boundaries in the \dirf100 and \dirf110 STGB systems.  Interestingly, the $\Sigma3$\plane112 GB has the same disorientation angle, tilt direction, and CSL value as the $\Sigma3$\plane111 GB, but has a very different behavior in terms of the binding behavior with He$_n$V clusters.  Also notice that the $\Sigma3$ GBs have the lowest and highest GB energies in the present study, which can explain their difference in binding energy behavior.  For instance, in a prior study with point defects in Fe \cite{Tsc2012a}, it was found that the mean formation energies (and consequently, binding energies) for each particular grain boundary was related to the grain boundary energy.  The other nine grain boundaries showed very similar maximum binding energies and binding energy behavior within the grain boundary region, which may be more typical of general high angle grain boundaries.  A similar spatial distribution of binding energies is observed for GBs in the \dirf100 symmetric tilt grain boundary system.

\begin{figure}[ht!]
  \centering
		  \includegraphics[width=\textwidth,angle=0]{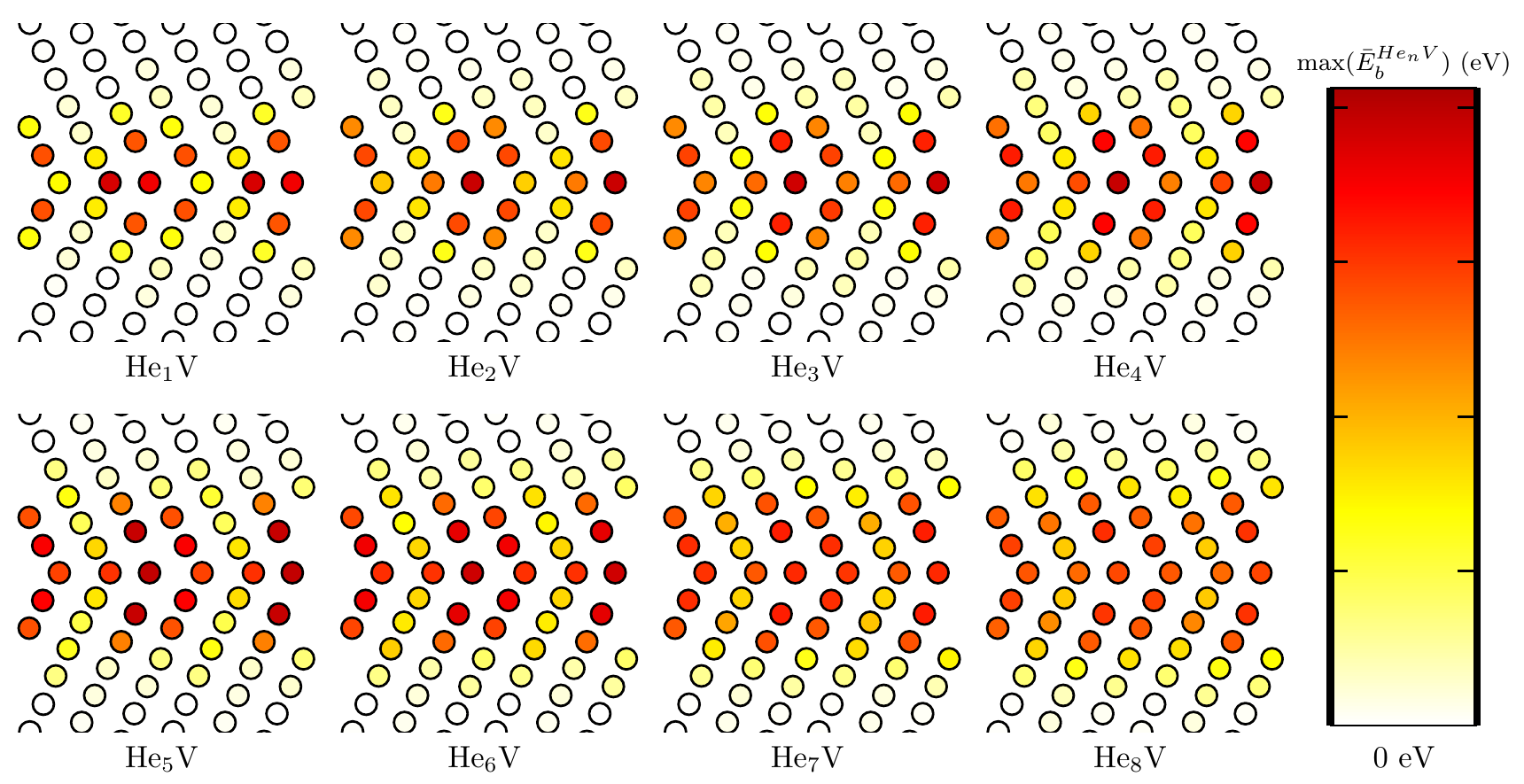}\\
\caption{ \label{figure3} The binding energies of the four different He$_n$V clusters in various sites for a representative high angle grain boundary ($\Sigma11$\plane113 GB).}
\end{figure}

The relative binding energy of the He$_n$V clusters with respect to the grain boundary structure is also of interest.  For instance, Figure \ref{figure3} shows the spatial distribution of binding energies for four He$_n$V clusters investigated in the present study for a representative high angle STGB: the $\Sigma11$\plane113 GB.  The contour bar for each He$_n$V cluster is scaled from 0 eV to the maximum binding energy among all 10 GBs for that particular He$_n$V cluster.  Therefore, the represented values are a relative measure with respect to the maximum binding energy for each He$_n$V cluster to facilitate comparison with respect to the number of He atoms $n$ in the cluster.  In many respects, the binding behavior is very similar between the different He$_n$V clusters.  The sites with the largest binding energies appears to stay consistent even as the number of He atoms in a monovacancy increase.  However, there are a few subtle differences that are also apparent in Figure \ref{figure3}.  First, the interaction length scale of the grain boundary increases as the number of He atoms in the He$_n$V cluster increases.  For instance, there are sites that are several layers away from the GB center with a relatively low relative binding energy for He$_1$V, but have a much higher relative binding energy for the He$_8$V cluster.  Second, focusing on the atoms within the grain boundary plane, there is less anisotropy in the relative binding energies as the number of He atoms in the He$_n$V cluster increases.  Hence, while the grain boundary structure may have a strong influence on the formation and binding energy of He clusters (or even solute atoms and impurities, for that matter) with only a few atoms, this effect may be less pronounced as the size of the cluster binding with the grain boundary increases.  In this case, the binding behavior may be more strongly influenced by the reduction in the grain boundary energy.  

\begin{table}
\centering
\caption{\label{table5} Linear correlation coefficients $R$ for the formation and binding energies of various He$_n$V clusters with each other.}
\begin{ruledtabular}
\begin{tabular}{@{}ccccccccc@{}}
Cluster & He$_1$V	& He$_2$V & He$_3$V & He$_4$V & He$_5$V & He$_6$V & He$_7$V & He$_8$V \\ 
\hline \\ [-2ex]
He$_1$V &  1.000 &  0.943 &  0.910 &  0.882 &  0.834 &  0.809 &  0.785 &  0.765 \\ 
He$_2$V &  -- &  1.000 &  0.976 &  0.949 &  0.911 &  0.887 &  0.859 &  0.837 \\ 
He$_3$V &  -- &  -- &  1.000 &  0.989 &  0.958 &  0.937 &  0.912 &  0.889 \\ 
He$_4$V &  -- &  -- &  -- &  1.000 &  0.983 &  0.968 &  0.946 &  0.925 \\ 
He$_5$V &  -- &  -- &  -- &  -- &  1.000 &  0.992 &  0.975 &  0.957 \\ 
He$_6$V &  -- &  -- &  -- &  -- &  -- &  1.000 &  0.990 &  0.978 \\ 
He$_7$V &  -- &  -- &  -- &  -- &  -- &  -- &  1.000 &  0.994 \\ 
He$_8$V &  -- &  -- &  -- &  -- &  -- &  -- &  -- &  1.000 
\end{tabular}
\end{ruledtabular}
\end{table}

\vspace{-10pt}
\subsection{Influence of Local GB Structure}

The local environment surrounding each atom changes as due to interactions with neighboring atoms, which in turn affects the cohesive energy and other per-atom properties.  In this subsection, we will analyze several metrics used to characterize the local environment surrounding each atom and compare with the previously calculated formation and binding energies for the different He$_n$V clusters.  In this work, six per-atom metrics are examined: the cohesive energy $E_{coh}$, the hydrostatic stress $\sigma_H$, the Voronoi volume $V_{Voro}$, the centrosymmetry parameter CS \cite{Kel1998}, the common neighbor analysis CNA \cite{Jon1988,Hon1987,Tsu2007}, and the coordination number CN.  The per-atom virial stress components were used to calculate $\sigma_H$ utilizing the Voronoi volume consistent with the bulk bcc $\alpha$-Fe lattice.  The per-atom Voronoi volume is also defined from a Voronoi tesselation of the space of the Fe atoms in a clean GB.  Previous studies have shown a good correlation between the Voronoi volume and the formation energies of He defects \cite{Zha2013,Tsc2014}.  Moreover, DFT studies have shown that increased Voronoi volumes lead to enhanced magnetic moments, higher local energies, and tensile stresses \cite{Bha2013}.  For CNA and CN, a cutoff distance of 3.5 \AA\ was used and eight nearest neighbors were used to calculate the CSP.  Over all GBs, CNA classifies the sites as 85.7\% bcc and 14.3\% unknown (866 sites total for 10 GBs).  The coordination numbers for these sites are: 1.2\% 11 CN, 2.5\% 12 CN, 4.0\% 13 CN, 91.0\% 14 CN (perfect bcc lattice), 0.6\% 15 CN, 0.7\% CN.  Hence, there are more undercoordinated sites than overcoordinated sites, as would be expected at the grain boundary (i.e., positive free volume).  Furthermore, the per-atom vacancy binding energies $E_b^v$ from Tschopp et al.~\cite{Tsc2011, Tsc2012a} will be compared as part of this analysis, which may be relevant since the He$_n$V clusters involve a monovacancy.  

The per-atom metrics were tabulated for all grain boundaries and compared with the He$_n$V clusters to find out which local environmental metrics are correlated.  The linear correlation coefficient $R$ is again used to compare the degree of correlation between the local metrics and the binding energies, where $R=1$ indicates a perfect positive correlation and $R=-1$ indicates a perfect negative correlation.  The results are shown in Table \ref{table6} along with results for interstitial He and He$_2$ \cite{Tsc2014}.  The binding energies are used instead of formation energies; realize that to change the correlation coefficient $R$ from binding energies $E_b$ to formation energies $E_f$, a factor of $-1$ can be applied to all $R$ values in Table \ref{table6}.  The CNA, CSP, and $E_{coh}$ have the highest positive correlation with the binding energies of the He$_n$V clusters (e.g., $R=0.85$, $R=0.77$, and $R=0.76$ for the He$_1$V cluster, respectively).  The hydrostatic stress $\sigma_H$ and CN have the lowest correlation with the He$_n$V cluster binding energies.  All metrics have a positive correlation with the binding energies of the He$_n$V clusters except for the coordination number, where undercoordinated atoms have higher binding energies and vice versa.  Several metrics have slightly higher correlations for $E_b$ of He$_1$V as well, indicating that as the degree of correlation decreases as the complexity of the He$_n$V cluster increases.  In particular, as the number of He atoms in the He$_n$V cluster increases, there is less of a linear correlation with the per-atom local environmental metrics.  This result is as expected.  The per-atom metrics refer to the unstrained environment prior to inserting the He$_n$V cluster; with an increasing number of He atoms in a monovacancy, the local environment is under a much larger volumetric strain than in the reference environment.  
 
\begin{table}
\centering
\caption{\label{table6} Linear correlation coefficients $R$ for the local per-atom metrics, $E_b^v$ and the He$_n$V binding energies.}
\begin{ruledtabular}
\begin{threeparttable}
\begin{tabular}{cccccccc}
Cluster $E_b$ & $E_b^v$ & $E_{coh}$ 	& $\sigma_H$ 	& $V_{Voro}$ 	& CSP 	& CNA 	& CN \\
\hline \\ [-1.5ex]
$E_b^{He_1V}$ &   0.67 &   0.76 &   0.20 &   0.63 &   0.77 &   0.85 &  -0.46 \\ 
$E_b^{He_2V}$ &   0.55 &   0.61 &   0.34 &   0.70 &   0.77 &   0.83 &  -0.34 \\ 
$E_b^{He_3V}$ &   0.52 &   0.57 &   0.35 &   0.69 &   0.75 &   0.81 &  -0.32 \\ 
$E_b^{He_4V}$ &   0.51 &   0.57 &   0.33 &   0.65 &   0.71 &   0.78 &  -0.31 \\ 
$E_b^{He_5V}$ &   0.49 &   0.55 &   0.30 &   0.60 &   0.67 &   0.75 &  -0.31 \\ 
$E_b^{He_6V}$ &   0.50 &   0.55 &   0.28 &   0.56 &   0.64 &   0.72 &  -0.31 \\ 
$E_b^{He_7V}$ &   0.51 &   0.55 &   0.24 &   0.52 &   0.60 &   0.69 &  -0.32 \\ 
$E_b^{He_8V}$ &   0.51 &   0.55 &   0.21 &   0.49 &   0.58 &   0.68 &  -0.31 \\  [0.5ex]
\hline \\ [-1.5ex]
$E_b^{HeInt}~$\tnote{\textdagger}  &  0.57 &   0.63 &   0.28 &   0.65 &   0.77 &   0.82 &  -0.36 \\ 
$E_b^{He_2Int}~$\tnote{\textdagger} &  0.56 &   0.63 &   0.28 &   0.63 &   0.75 &   0.80 &  -0.34 \\
\end{tabular}
        \begin{tablenotes}
		\scriptsize
		\item [\textdagger] Interstitial He and He$_2$ binding energies are taken from Ref.~\onlinecite{Tsc2014}.
        \end{tablenotes}
\end{threeparttable}
\end{ruledtabular}
\end{table}

\vspace{-10pt}
\subsection{Statistical GB Description}

\begin{figure}[b!]
  \centering
	\includegraphics[width=0.95\columnwidth,angle=0]{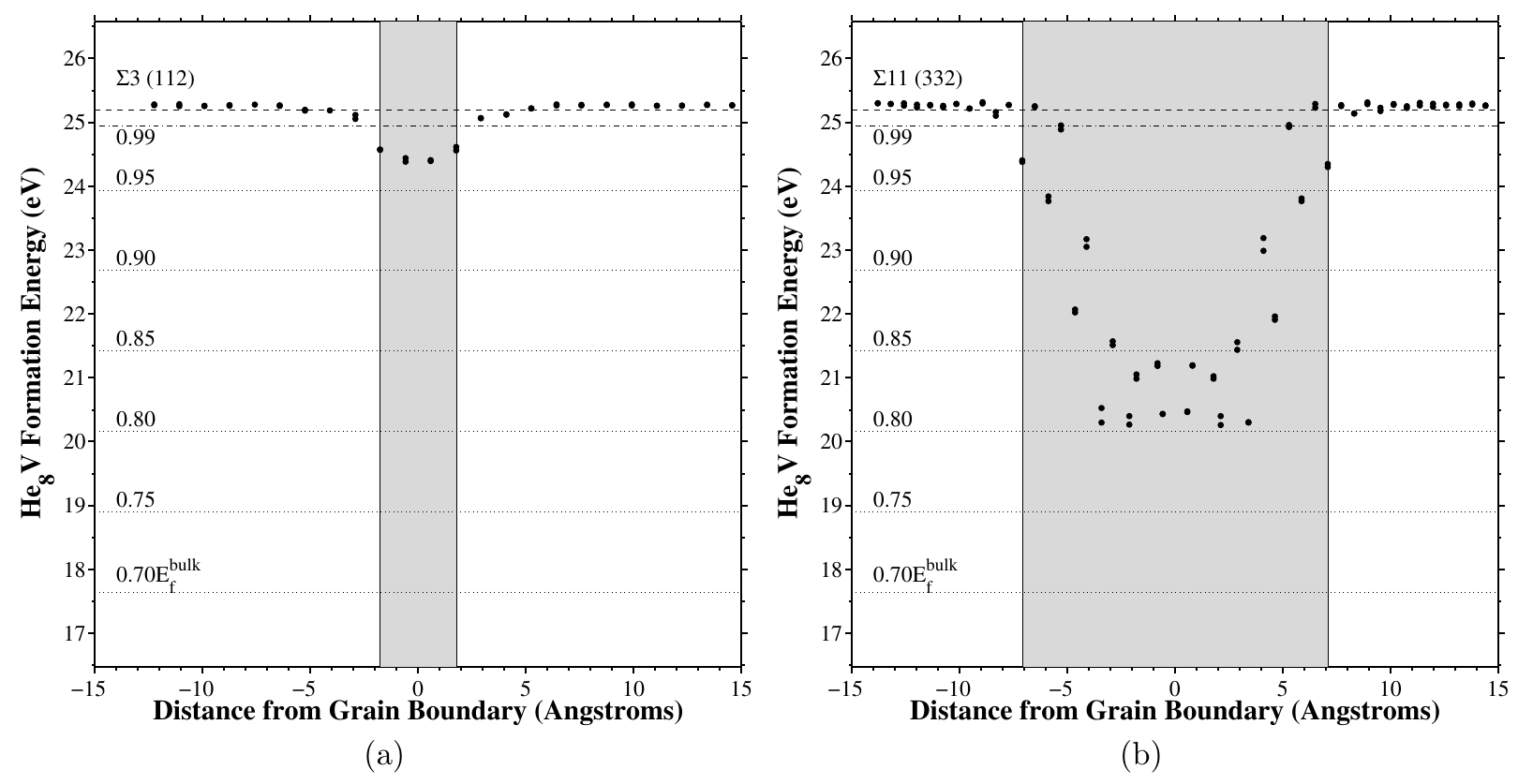} \\
\caption{ \label{figure4} The formation energies of He$_8$V cluster in various sites plotted against the distance from the grain boundary for the (a) $\Sigma3$\plane112 and (b) $\Sigma11$\plane332 GBs.  }
\end{figure}

The formation energies of the eight He$_n$V clusters can be plotted against the distance from the grain boundary to quantify the evolution of the formation energies (and binding energies) near the GB and to quantify the length scale associated with the He$_n$V clusters.  Figure \ref{figure4} is an example of one such plot for the He$_8$V cluster at various sites at the  $\Sigma3$\plane112 GB and $\Sigma11$\plane332 GB.  In this plot, the formation energies $E_f^{He_8V\alpha}$ of each of the twenty different He$_8$V instantiations was first calculated for each site $\alpha$ and the mean formation energy  $\bar{E}_f^{He_8V\alpha}$ was subsequently calculated from these values.  Next, a grain boundary region was defined to compare the different He$_n$V clusters and the different GBs examined in the current work. This grain boundary region was subsequently used to quantify three parameters: the segregation length scale $l_{GB}$, the mean binding energy $\tilde{E}_b^{mean}$, and the maximum binding energy $\tilde{E}_b^{max}$.  The length scale parameter $l_{GB}$ is calculated by first defining a subset $\beta$ of all sites $\alpha$ based on deviation of formation energies from the bulk formation energy, i.e., 
\begin{equation}
  \label{eq:eq5}
	\beta = \left\{\alpha|\bar{E}^{He_nV^\alpha}_{f}\le0.99E_f^{bulk} \right\}, \\
\end{equation}
and then calculating the bounds of the grain boundary affected region, i.e., 
\begin{equation}
\label{eq:eq6}
\begin{aligned}
	x_{min} &= \textrm{min} \left(\mathbf{x}^\beta\right), \\	
	x_{max} &=\textrm{max} \left(\mathbf{x}^\beta\right) \\
	l_{GB} &= x_{max}-x_{min},	
\end{aligned}
\end{equation}
\noindent where $\beta$ is a subset of all sites $\alpha$ where the above condition is met and $\mathbf{x}^\beta$ is the vector containing the coordinates of all sites $\beta$ in the direction perpendicular to the grain boundary plane.  In the subsequent plots and analysis, the coordinate for the grain boundary plane was shifted such that $x=0$ is the center of the grain boundary and there is an equal distance to the bounds of the grain boundary affected region, $x_{min}$ and $x_{max}$. Then, to calculate the average binding properties of the grain boundary affected region, a subset $\gamma$ that contains all sites within the grain boundary affected region is defined, i.e., 
\begin{equation}
  \label{eq:eq7}
	\gamma = \left\{\alpha|x_{min}\ge\mathbf{x}^\alpha\ge{x}_{max}. \right\} \\
\end{equation}
\noindent Note that $\gamma$ and $\beta$ are not necessarily equivalent sets since $\gamma$ includes all sites within the grain boundary region and $\beta$ included only those sites with formation energies that were different from the bulk ($\le0.99E_f^{bulk}$).  The mean and maximum binding energies for a particular He$_n$V cluster are now given by
\begin{equation}
  \label{eq:eq8}
	\tilde{E}^{mean}_{b}=\frac{1}{n_\gamma}\sum_{\gamma=1}^{n_\gamma}\bar{E}^{He_nV^\gamma}_{b},	
\end{equation}
and
\begin{equation}
  \label{eq:eq9}
	\tilde{E}^{max}_{b}=\mathrm{max}\left(\bar{E}^{He_nV^\gamma}_{b}\right),	
\end{equation}
\noindent where the summation sign in Eq.~\ref{eq:eq6} operates over the number of $\gamma$ sites $n_\gamma$ within the specified length scale $l_{GB}$.  First, the difference between the mean formation energy at $>$10 \AA\ and the formation energy calculated in a 2000 atom single crystal unit cell was calculated for each boundary and any bias detected was subsequently removed.   Prior simulations to test for convergence of formation energies as a function of  simulation cell size show that this bias was associated with the simulation cell size.  The simulation cell sizes given in Table \ref{table1} produced a bias on the order of $0.01E_f^{bulk}$ or less.  The GB-affected region identified in Eq.~\ref{eq:eq6} is shaded light gray in Fig.~\ref{figure4} and is bounded by the coordinates $x_{min}$ and $x_{max}$, which corresponds to sites where the formation energies first deviate by more than $0.01E_f^{bulk}$ from the bulk formation energy $E_f^{bulk}$ ($0.99E_f^{bulk}$ is one of the dashed lines in Figure \ref{figure4}).  The minimum length that encompasses all these GB sites is $l_{GB}$.  Next, the formation energies were converted to binding energies for each He$_n$V cluster/GB combination to compare the energy gained by each defect segregating to the boundary as opposed to the bulk lattice.  Both the mean and maximum binding energies ($\tilde{E}_b^{mean}$ and $\tilde{E}_b^{max}$ in Eqs.~\ref{eq:eq8} and \ref{eq:eq9}, respectively) for this region are then calculated.  To illustrate the percent difference from the bulk formation energy, increments of $0.05E_f^{bulk}$, or 5\% of the bulk formation energy, are indicated by dotted lines in Figure \ref{figure4}.  For instance, in the $\Sigma11$\plane332 GB, the maximum binding energy is $\approx$20\% of the bulk formation energy (i.e., $\approx0.80E_f^{bulk}$) and lies towards the center of the grain boundary region.  This technique for identifying three parameters for He segregation was subsequently applied to all 10 GBs for all He$_n$V clusters.

The length scale $l_{GB}$, mean binding energy $\tilde{E}_b^{mean}$, and maximum binding energy $\tilde{E}_b^{max}$ for all eight He$_n$V clusters to the ten grain boundaries examined in the current study is tabulated in Tables \ref{table7}, \ref{table8}, and \ref{table9}, respectively.  It is immediately apparent that the $\Sigma3$\plane112 twin boundary has a smaller length scale and smaller binding energies than the other boundaries.  As observed in Table \ref{table1}, this boundary has both the lowest energy and lowest free volume, which supports that these macroscale GB parameters may indicate lower binding energies with He defects (e.g., as suggested by Kurtz et al.~\cite{Kur2004}).  The other boundaries have very similar length scales (typically between 8 \AA\ to 12 \AA) and binding energies (sensitive to He defect type), with a few instances where one boundary has an interaction length scale or binding energies different from the rest (e.g., the $\Sigma13$\plane320 GB consistently has the largest GB-affected region).  As a general trend, the length scale $l_{GB}$ increases with increasing number of He atoms $n$ for the He$_n$V cluster types (Table \ref{table7}).  Moreover, both the mean and maximum binding energies (Tables \ref{table8} and \ref{table9}) also increase with increasing $n$. 

\begin{table}
\centering
\caption{\label{table7} Length scale $l_{GB}$ for various He$_n$V clusters for the ten grain boundaries (in \AA)}
\begin{ruledtabular}
\begin{tabular}{ccccccccc}
GB Type & He$_1$V & He$_2$V & He$_3$V & He$_4$V & He$_5$V & He$_6$V & He$_7$V & He$_8$V \\
\hline \\ [-1.5ex]
     $\Sigma3$ \planef111 &   7.0 &   8.5 &   8.5 &   8.5 &  13.5 &  13.5 &  13.5 &  13.5  \\ 
     $\Sigma3$ \planef112 &   3.5 &   3.5 &   3.5 &   3.5 &   3.5 &   3.5 &   3.5 &   3.5  \\ 
     $\Sigma5$ \planef210 &   8.0 &   9.3 &  10.5 &  10.5 &  11.2 &  12.4 &  12.4 &  14.4  \\ 
     $\Sigma5$ \planef310 &   7.5 &   5.8 &   5.8 &   5.8 &   5.8 &   6.6 &   7.5 &   9.4  \\ 
     $\Sigma9$ \planef221 &   8.3 &  10.6 &  10.6 &  10.6 &  10.6 &  12.6 &  12.6 &  15.4  \\ 
     $\Sigma9$ \planef114 &   6.4 &   9.1 &  10.4 &  10.4 &  10.4 &  11.8 &  11.8 &  13.1  \\ 
    $\Sigma11$ \planef113 &   8.0 &  10.2 &  10.6 &  10.6 &  11.5 &  11.5 &  11.5 &  14.0  \\ 
    $\Sigma11$ \planef332 &   8.2 &   9.3 &   9.3 &  11.7 &  11.7 &  14.2 &  14.2 &  14.2  \\ 
    $\Sigma13$ \planef510 &   7.5 &   8.7 &   9.2 &   9.2 &  10.3 &  10.9 &  10.9 &  10.9  \\ 
    $\Sigma13$ \planef320 &  13.1 &  16.8 &  17.2 &  18.9 &  18.9 &  18.9 &  18.9 &  20.9 
\end{tabular}
\end{ruledtabular}
\end{table}

\begin{table}
\centering
\caption{\label{table8} Mean binding energy $\tilde{E}_b^{mean}$ for various He$_n$V clusters for the ten grain boundaries (in eV)}
\begin{ruledtabular}
\begin{tabular}{ccccccccc}
GB Type & He$_1$V & He$_2$V & He$_3$V & He$_4$V & He$_5$V & He$_6$V & He$_7$V & He$_8$V \\
\hline \\ [-1.5ex]
     $\Sigma3$ \planef111 &  0.41 &  1.07 &  1.91 &  2.49 &  2.34 &  2.77 &  3.63 &  4.38  \\ 
     $\Sigma3$ \planef112 &  0.07 &  0.10 &  0.19 &  0.26 &  0.37 &  0.46 &  0.43 &  0.58  \\ 
     $\Sigma5$ \planef210 &  0.37 &  0.81 &  1.22 &  1.62 &  2.28 &  2.63 &  3.25 &  3.35  \\ 
     $\Sigma5$ \planef310 &  0.33 &  1.00 &  1.64 &  2.19 &  3.02 &  3.04 &  3.82 &  3.87  \\ 
     $\Sigma9$ \planef221 &  0.35 &  0.77 &  1.21 &  1.59 &  2.36 &  2.62 &  3.29 &  3.31  \\ 
     $\Sigma9$ \planef114 &  0.45 &  0.83 &  1.18 &  1.59 &  2.26 &  2.53 &  3.36 &  3.81  \\ 
    $\Sigma11$ \planef113 &  0.31 &  0.63 &  1.02 &  1.44 &  1.95 &  2.43 &  3.19 &  3.44  \\ 
    $\Sigma11$ \planef332 &  0.29 &  0.68 &  1.09 &  1.16 &  1.81 &  1.88 &  2.36 &  3.00  \\ 
    $\Sigma13$ \planef510 &  0.35 &  0.74 &  1.11 &  1.51 &  2.04 &  2.28 &  2.92 &  3.67  \\ 
    $\Sigma13$ \planef320 &  0.19 &  0.37 &  0.67 &  0.82 &  1.21 &  1.48 &  1.87 &  2.13 
\end{tabular}
\end{ruledtabular}
\end{table}

\begin{table}
\centering
\caption{\label{table9} Maximum binding energy $\tilde{E}_b^{max}$ for various He$_n$V clusters for the ten grain boundaries (in eV)}
\begin{ruledtabular}
\begin{tabular}{ccccccccc}
GB Type & He$_1$V & He$_2$V & He$_3$V & He$_4$V & He$_5$V & He$_6$V & He$_7$V & He$_8$V \\
\hline \\ [-1.5ex]
     $\Sigma3$ \planef111 &  0.79 &  1.49 &  2.60 &  2.98 &  3.91 &  4.36 &  6.05 &  7.29  \\ 
     $\Sigma3$ \planef112 &  0.13 &  0.17 &  0.30 &  0.41 &  0.54 &  0.65 &  0.68 &  0.89  \\ 
     $\Sigma5$ \planef210 &  0.75 &  1.69 &  2.44 &  2.99 &  4.04 &  4.83 &  5.95 &  6.90  \\ 
     $\Sigma5$ \planef310 &  0.68 &  1.35 &  2.27 &  2.77 &  3.30 &  3.95 &  4.74 &  5.55  \\ 
     $\Sigma9$ \planef221 &  0.79 &  1.56 &  2.42 &  2.83 &  4.14 &  4.76 &  5.51 &  6.59  \\ 
     $\Sigma9$ \planef114 &  0.71 &  1.52 &  2.38 &  2.98 &  3.88 &  4.40 &  5.73 &  6.73  \\ 
    $\Sigma11$ \planef113 &  0.75 &  1.75 &  2.71 &  3.44 &  4.29 &  4.90 &  5.41 &  6.24  \\ 
    $\Sigma11$ \planef332 &  0.49 &  1.22 &  1.91 &  2.24 &  3.00 &  3.31 &  4.09 &  5.01  \\ 
    $\Sigma13$ \planef510 &  0.79 &  1.86 &  2.91 &  3.64 &  4.43 &  4.98 &  6.24 &  7.36  \\ 
    $\Sigma13$ \planef320 &  0.83 &  1.78 &  2.64 &  3.50 &  4.49 &  5.25 &  6.91 &  8.38 
\end{tabular}
\end{ruledtabular}
\end{table}

\vspace{-10pt}
\subsection{Atomistically-informed Model for Binding Energy Distribution}

\begin{figure}[b!]
  \centering
	\includegraphics[width=0.975\columnwidth,angle=0]{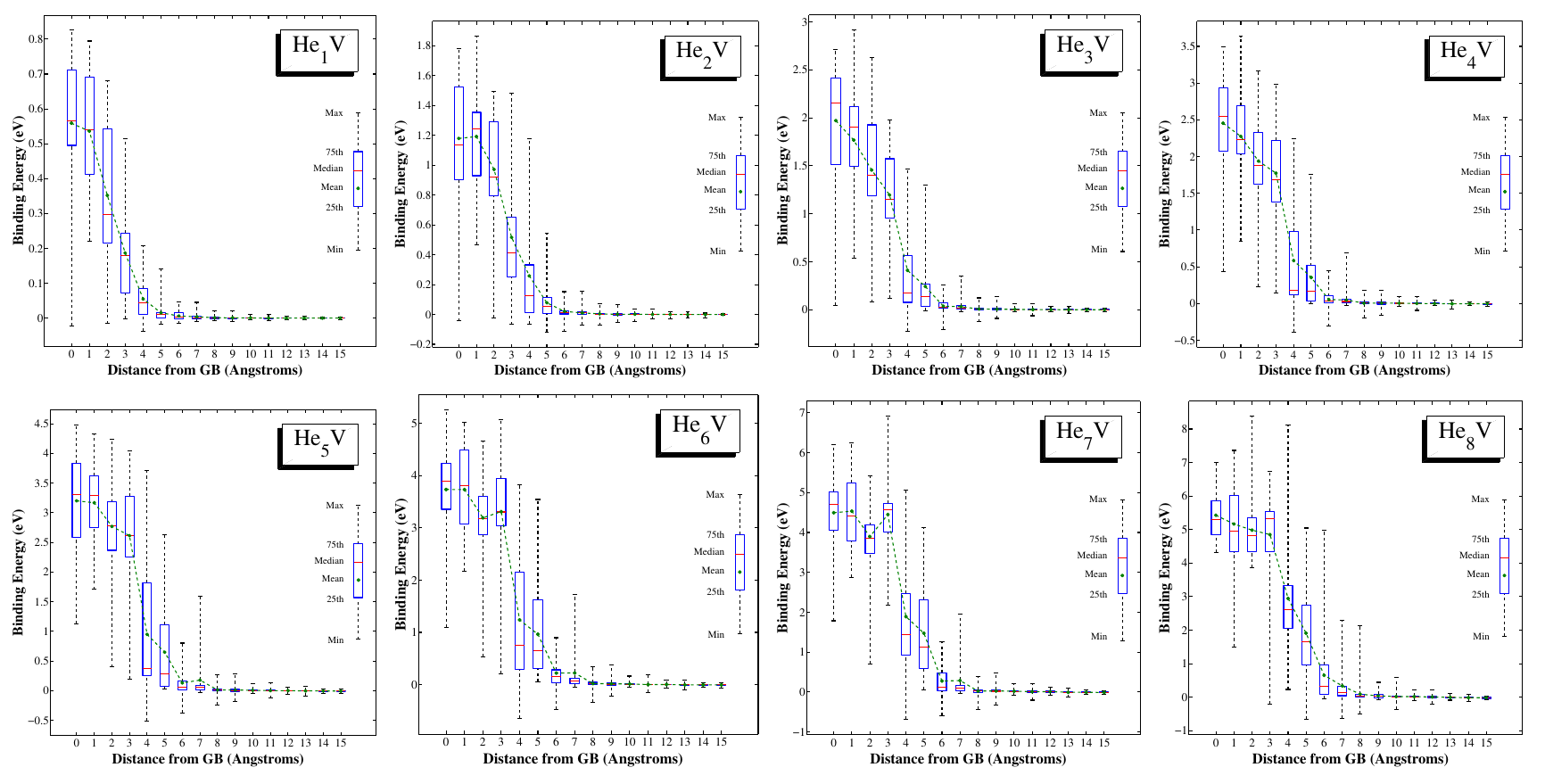}  
\caption{ \label{figure5} Boxplots of binding energy as a function of distance from the grain boundary for the nine representative GBs (excluding the $\Sigma3$\plane112) for various He$_n$V clusters: (top, left to right) He$_1$V, He$_2$V, He$_3$V, He$_4$V,  (bottom, left to right) He$_5$V, He$_6$V, He$_7$V, He$_8$V.  The data is divided into 1 \AA\ bins, and a boxplot is made for each bin. The red lines are medians, the blue box ends are the first and third quartiles, and the black whisker ends are minimum and maximum values.  The mean binding energy is also plotted in green.  }
\end{figure}

The change in binding energies of He$_n$V clusters as a function of distance from the GB center can also be analyzed by binning the energies and calculating the statistics associated with each bin (Figure \ref{figure5}).  Due to the symmetric nature of the GB formation and binding energies as a function of distance (e.g., Figure \ref{figure4}), the absolute value of the distance from the GB center was used to provide more data points for the statistical analysis.  Furthermore, the energies are split into 1 \AA\ bins to characterize the distributions and compute statistics for sites at a given distance from the GB.  For example, the 0 \AA\ bin would contain all binding energies for sites within $-0.5$ \AA\ to $+0.5$ \AA\ from the GB center and then several statistics are calculated from these binding energy distributions.  A boxplot (Figure \ref{figure5}) is used to represent the binding energy distribution in each bin, i.e., the minimum, 25\% percentile, median, 75\% percentile, and maximum binding energies.  In the boxplot, the red line in the box is the median while the  bottom and top edges of the blue boxes represent the 25\% and 75\% quartiles (as shown to the right of each plot). The whiskers extending from the boxes cover the remainder of the range of energies for each bin, and the ends of the whiskers denote the maximum and minimum values of the binding energies for each bin.  The mean value of the binding energies in each bin is also plotted in green.  Boxplots can be very useful for visualizing any asymmetry in the distribution of energies.

The box plots in Figure \ref{figure5} encompass all the binding energy data for all He$_n$V clusters from the nine representative GBs (excluding the $\Sigma3$\plane112).  The mean binding energy is largest for sites close to the GB (0 and 1 \AA\ bins), as shown in Figure \ref{figure5}, and it approaches the normalized bulk value of zero as sites are located farther from the boundary.  The mean and median values of the binding energies trend together.  Similar to the trends found in Table \ref{table7}, there is a definite length scale associated with He$_n$V clusters binding to the grain boundary that is on the order of 4--6 \AA\ from the GB center.  For binding energies within the GB region, the distribution of binding energy is slightly skewed (for a symmetric distribution, the red line lies exactly in the middle of the box) with a large degree of variability, as can be seen from both the difference between the minimum and maximum values as well as the magnitude of the interquartile range (height of the boxes, denoting the binding energies associated with the 25\% and 75\% percentiles)).    At distances $>$7 \AA, the binding energy distribution trends towards zero, indicating that the overwhelming majority of atomic sites display a binding energy similar to the bulk value.  

A model was formulated to capture the formation energy and binding energy distribution evolution as a function of distance.  As the general behavior is symmetric about zero and approaches zero at large distances, an exponential formulation is used to describe the change in the mean $\mu$ and standard deviation $\sigma$ of the binding energy distribution as a function of distance from the grain boundary $x$:
\begin{equation}
  \label{eq:eq10}
	\tilde{E}^{\{\mu,\sigma\}}_{b}\left(x\right)=c_1\,\mathrm{exp}\left(-\left(\frac{x}{c_2}\right)^{c_3}\right) ,	
\end{equation}
\noindent where $c_1$, $c_2$, and $c_3$ are model constants.  Notice that $c_3=2$ will give a form similar to a normal distribution.  Based on an initial nonlinear fit of the data, $c_3=2$ was chosen for $ \tilde{E}^{\mu}_{b}\left(x\right)$ and $c_3=4$ was chosen for $\tilde{E}^{\sigma}_{b}\left(x\right)$; these values were kept constant to better assess the change in the constant $c_2$.  Then, a fit of $c_1$ and $c_2$ to the mean (in Figure \ref{figure5}) and standard deviation (not shown) curves leads to the values given in Table \ref{table10} for all He$_n$V clusters.  Interestingly, the constant $c_1$ is related to the mean and maximum binding energy ($\tilde{E}_b^{mean}$ and $\tilde{E}_b^{max}$ in Tables \ref{table8} and \ref{table9}, respectively) and the constant $c_2$ is related to the length scale $l_{GB}$ (Table \ref{table7}).  The root mean square error (RMSE) in Table \ref{table10} increases with an increasing number of He atoms $n$ in the monovacancy and roughly correlates with the constant $c_1$ (the ratio $\mathrm{RMSE}/c_1$ is shown in parenthesis).  The coefficient of determination $R^2$ for the model given by Equation \ref{eq:eq10} tends to decrease with increasing $n$, but most $R^2$ values are above $0.90$.  While additional terms can be added to improve the model agreement, this simple model does show the same trends with  $c_1$ and $c_2$ as was shown earlier with $\tilde{E}_b^{max}$ and $l_{GB}$; the length scale increases with increasing $n$ as does the maximum binding energy.  Additionally, the present model can be modified to obtain the variation in formation energies for He$_n$V clusters, i.e., $\tilde{E}^{\mu}_{f}\left(x\right)=E_f^{bulk}-\tilde{E}^{\mu}_{b}\left(x\right)$ and $\tilde{E}^{\sigma}_{f}\left(x\right)=\tilde{E}^{\sigma}_{b}\left(x\right)$, where $\tilde{E}^{\mu,}_{f}\left(x\right)$ and $\tilde{E}^{\sigma}_{f}\left(x\right)$ are the mean and standard deviation of the formation energy distribution with distance from the GB.  In this manner, the present model can utilize atomistic and/or DFT binding/formation energies to analytically describe the He$_n$V energetics as a function of distance from the grain boundary.  The aforementioned molecular simulations, analysis, and the resulting analytical formulation can be useful for incorporating the energetics and length scales of grain boundary segregation into higher order models.

\begin{table}
\centering
\caption{\label{table10} Parameterization for binding energy as a function of distance.}
\begin{ruledtabular}
\begin{tabular}{lcccccccc}
 \\ [-2.5ex]
\multirow{2}*{\begin{minipage}{0.5in}He$_n$V Cluster\end{minipage}} &  \multicolumn{4}{c}{$\tilde{E}^{\mu}_{b}\left(x\right)$}
&  \multicolumn{4}{c}{$\tilde{E}^{\sigma}_{b}\left(x\right)$}  \\ [0.5ex]
\cline{2-9} \\ [-1.5ex]
& $c_1$ (eV) & $c_2$ (\AA) & RMSE (eV)  & $R^2$ & $c_1$ (eV) & $c_2$ (\AA)  & RMSE (eV)  & $R^2$  \\ [0.5ex]
\hline \\ [-1.5ex]
He$_1$V &   0.58 &   2.78 &  0.011 (0.02) &  0.997  &   0.20 &   4.07 &  0.010 (0.05) &  0.986 \\
He$_2$V &   1.26 &   3.26 &  0.041 (0.03) &  0.992  &   0.39 &   5.10 &  0.032 (0.08) &  0.965 \\
He$_3$V &   1.98 &   3.56 &  0.078 (0.04) &  0.989  &   0.59 &   5.43 &  0.046 (0.08) &  0.969 \\
He$_4$V &   2.53 &   3.78 &  0.143 (0.06) &  0.978  &   0.70 &   5.78 &  0.058 (0.08) &  0.966 \\
He$_5$V &   3.42 &   4.05 &  0.216 (0.06) &  0.973  &   0.88 &   6.70 &  0.107 (0.12) &  0.932 \\
He$_6$V &   4.00 &   4.24 &  0.285 (0.07) &  0.967  &   1.00 &   6.93 &  0.136 (0.14) &  0.915 \\
He$_7$V &   4.87 &   4.52 &  0.413 (0.08) &  0.955  &   1.07 &   7.22 &  0.165 (0.15) &  0.893 \\
He$_8$V &   5.77 &   4.75 &  0.347 (0.06) &  0.977  &   1.26 &   7.79 &  0.236 (0.19) &  0.851 \\
\end{tabular}
\end{ruledtabular}
\end{table}

\vspace{-10pt}
\section{Conclusions}

The formation/binding energetics and length scales associated with the interaction between He$_n$V clusters and grain boundaries in bcc $\alpha$-Fe was explored.  Ten different low $\Sigma$ grain boundaries from the \dirf100 and \dirf110 symmetric tilt grain boundary systems were used (Table \ref{table1}) along with an Fe--He interatomic potential fit to ab initio calculations \cite{Gao2011} (Table \ref{table2}).  In this work, we then calculated formation/binding energies for 1--8 He atoms in a monovacancy (He$_1$V--He$_8$V) at all potential grain boundary sites within 15 \AA\ of the boundary (122106 simulations total).  To account for the potential variability in He$_n$V cluster configurations, 20 different random starting positions for the He$_n$ atoms about each monovacancy were simulated (Figure \ref{figure1}).  The present results provide detailed information about the interaction energies and length scales of 1--8 He atoms with grain boundaries for the structures examined.  The following conclusions can be drawn about this work:

\begin{itemize}
	\item The local atomic structure and spatial location within the boundary affects the magnitude of the formation/binding energies for all He$_n$V clusters (Figs.~\ref{figure2} and \ref{figure3}).  In general, grain boundary sites have much lower formation energies and higher binding energies than in the bulk, indicating an energetic driving force for He$_n$V clusters to reside in grain boundary sites.  This GB affected region visibly extends several planes from the GB center.  The maximum formation energy and binding energy (to the GB) for the He$_n$V clusters increases with an increasing number of He atoms in the monovacancy (Table \ref{table4}).  Furthermore, the $\Sigma3$\plane112 GB has significantly lower binding energies than all other GBs in this study, in agreement with previous results for interstitial He and He$_2$ \cite{Tsc2014}.
	\item The relative binding energy behavior was examined with respect to grain boundary structure (e.g., Figure \ref{figure3}).  As $n$ increases, the length scale of the GB-affected region increases and there is less variability in the binding energies with GB structure.  In fact, while the binding energy behavior between the He$_n$V clusters is linearly correlated in a positive sense, this correlation decreases as the number of He atoms between two He$_n$V clusters increases.  Metrics for quantifying or classifying the local structure of each atom site were also compared to the formation/binding energies of He clusters.  Trends in per-atom metrics with the He$_n$V cluster energies were tabulated in the form of linear correlation coefficients (Table \ref{table7}).  While common neighbor analysis (CNA), centrosymmetry parameter (CSP), cohesive energy $E_{coh}$, vacancy binding energy $E_b^v$, and Voronoi volume $V_{Voro}$ all positively correlated with He$_n$ cluster binding energies (with $R$ as high as 0.85, respectively), this correlation generally decreased with increasing $n$ and was never highly correlated ($R>0.90$).
	\item The change in formation and binding energies as a function of spatial position (Figure \ref{figure4}) was used to identify a GB affected region and to assess a corresponding length scale, mean binding energy, and maximum binding energy for this region (Tables \ref{table8}, \ref{table9}, \& \ref{table10}).  These plots were additionally reduced via symmetry about the GB plane (Figure \ref{figure5}) to show the evolution of the binding energy distribution as a function of distance from the GB plane for the various He$_n$V clusters.  Based on these results, we formulated a model to capture the evolution of the formation and binding energy of He$_n$V clusters as a function of distance from the GB center, utilizing only constants related to the maximum binding energy and the length scale.
\end{itemize}

This work significantly enhances our understanding of the energetics involved with how the grain boundary structure interacts with He$_n$V clusters and how ultimately this affects He (re-)combination and embrittlement near grain boundaries in polycrystalline steels.

\vspace{-10pt}
\section*{Acknowledgments}

This work is supported in part by the U.S. Army Research Laboratory (ARL).  F.G. is grateful for the support by the US Department of Energy, Office of Fusion Energy Science, under Contract DE-AC06-76RLO 1830.  The authors would like to acknowledge the support and discussions with Dr.~Xin Sun and Dr.~Moe Khaleel at Pacific Northwest National Laboratory.  Last, the authors would like to acknowledge Ms.~Joanna Sun, a high school student support by the Alternate Sponsored Fellowship (ASF) at PNNL, for her contributions to this work.

\bibliographystyle{unsrt}

\end{document}